\documentclass[prb,bibnotes,twocolumn,floatfix,showpacs]{revtex4}
\usepackage{graphicx}
\usepackage[tbtags]{amsmath}

\newcommand{\beq}{\begin{equation}}
\newcommand{\eeq}{\end{equation}}
\newcommand{\eeql}[1]{\label{#1} \end{equation}}
\newcommand{\phs}{{\vphantom{*}}}
\newcommand{\ij}{i\kern -0.08em j}
\def\hn{\mskip-0.5\thinmuskip}

\begin{document}
\title{Josephson-phase qubit without tunneling}
\author{M.H.S. Amin}
\email{amin@dwavesys.com}
\author{A.Yu.\ Smirnov}
\email{anatoly@dwavesys.com}
\author{Alec \surname{Maassen van den Brink}}
\thanks{Corresponding author;\\ electronic address: \texttt{alec@dwavesys.com}}
\affiliation{D-Wave Systems Inc., 320-1985 W. Broadway,
Vancouver, B.C., V6J 4Y3, Canada}
\date{\today}

\begin{abstract}
We show that a complete set of one-bit gates can be realized by coupling the two logical states of a phase qubit to a third level (at higher energy) using microwave pulses. Thus, one can achieve coherent control without invoking any tunneling between the qubit levels. We propose two implementations, using rf-SQUIDs and $d$-wave Josephson junctions.
\end{abstract}

\pacs{03.67.Lx
, 85.25.Cp
, 85.25.Dq
, 74.72.-h
}
\maketitle

In the field of Josephson qubits,\cite{chqubit} phase qubits enjoy continued attention. This is partly due to their tolerance to decohering background-charge noise compared to charge qubits. Most phase-based designs rely on the tunnel splitting $\Delta$ to flip the state, i.e., to perform a $\sigma_{\hn x}$ operation. This has several disadvantages. First, $\Delta$ is exponentially sensitive to the device parameters. This makes manufacturing spread especially severe, hampering scalability. Second, it is hard to stop the evolution, so one may need to e.g.\ refocus.\cite{refocus} One can in principle switch off $\Delta$ using a compound (Bloch-transistor) junction, but this considerably increases the parameter sensitivity even further.\cite{FA} Also, for many systems $\Delta$ is too small to be useful or even observable. Conversely, current-biased ``large-junction'' qubits\cite{martinis,yu} avoid the reliance on tunneling at the price of a large spacing between the logical levels, leading to a strong always-on $\sigma_{\hn z}$ evolution.

Recently, in Ref.~\onlinecite{han} it has been proposed to flip the state of a qubit by two consecutive microwave pulses. The first pulse excites the qubit from, say, $|0\rangle$ to a higher\cite{other} auxiliary state $|2\rangle$ through a Rabi oscillation. The next takes the qubit back to the logical space, but now to $|1\rangle$, addressing the first disadvantage above. However, this pulse sequence would carry $|1\rangle$ to $|2\rangle$ instead of the desired $|0\rangle$; \emph{a fortiori}, it thus does not map a general (superposition) qubit state to another, hence is not a valid gate operation. Even if this would be remedied [by, e.g., preceding (following up) the sequence with an extra $|1\rangle\leftrightarrow|2\rangle$ ($|0\rangle\leftrightarrow|2\rangle$) pulse], the method's state selectivity relies on a bias between $|0\rangle$ and $|1\rangle$, so the second disadvantage is overcome at best partially (the bias can be removed during idle periods, but not during gate action); also, the inflexible restriction to bit-flip gates only remains.

Simultaneously, Ref.~\onlinecite{BHK} has given a largely\cite{H2} correct proposal of using an auxiliary state to implement some gates for a different class of qubits. The coupling to the third state does not involve microwaves, and the resulting lack of tunability seems to limit the proposal to a discrete set of gates. In this paper, we resolve the abovementioned problems by showing that a general quantum gate \emph{can} be realized with Rabi pulses alone, without using tunneling.

\begin{figure}
\includegraphics[width=3in]{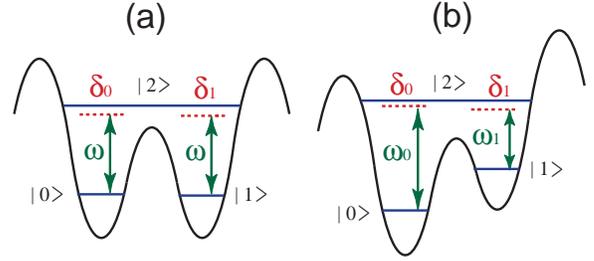}
\caption{Gate operation by coupling the two logical states to a third level with microwaves. (a)~(Near-)degenerate states and one pulse frequency. (b)~Biased states and two frequencies.}
\label{fig1}
\end{figure}

Consider a general system with a bistable potential (Fig.~\ref{fig1}). The lowest levels in the left and right wells are taken as the logical states $|0\rangle$,~$|1\rangle$. Unlike most other phase-qubit designs, we choose our parameters so as to make $\Delta$ smaller than all relevant energy scales, in particular the decoherence rate: $\Delta\ll1/\tau_\varphi$ ($\hbar=1$). Then, one can consider $|0\rangle$, $|1\rangle$ as energy eigenstates.

We induce transitions to a higher state $|2\rangle$ by applying microwaves with frequencies near the energy differences $E_2-E_{0,1}$ (Fig.~\ref{fig1}). The system then undergoes Rabi oscillations, starting from the logical space. After half a Rabi period ($t_\mathrm{c}=\pi/\Omega_\mathrm{R}$), the probability of finding the system in $|2\rangle$ will be zero again. The qubit wave function, however, will in general have changed: if, e.g., the system starts from $|0\rangle$, it will end up in a superposition. Thus, a matrix element has been created between $|0\rangle$ and $|1\rangle$, equivalent to a $\sigma_{\hn x}$ term in the reduced Hamiltonian.

More quantitatively, let us write the Hamiltonian as
\begin{gather}
  H = H_0 + V(t)\;,\label{H}\\
  H_0=E_0|0\rangle\langle0|+E_1|1\rangle\langle1|\label{H0}\;,
\end{gather}
where $H_0$ accounts for the uncoupled Josephson junction (with $E_2\equiv0$ for convenience) and $V(t)$ for the microwaves. This simple model captures the physics outlined above in two regimes. \emph{Regime~(a)} corresponds to near-degenerate logical levels and one external frequency,
\begin{subequations}
\beq
  V(t)=Ve^{-i\omega t}+ \text{h.c.}
\eeq
and $|\delta_j|\lesssim|V|$, with the detunings $\delta_j\equiv\omega+E_j$ ($j=0,1$) and $|V|$ the size of a typical matrix element of $V$ (Fig.~\ref{fig1}a). On the other hand, in \emph{regime~(b)} the logical levels are well-separated, $|E_1-E_0|\gg|V|$, and each is coupled to $|2\rangle$ by its own frequency:
\beq
  V(t) = V_0 e^{-i \omega_0t} + V_1 e^{-i\omega_1t} + \text{h.c.}
\eeql{Vt}
\end{subequations}%
Again $|\delta_j|\lesssim|V|$, with now $\delta_j\equiv\omega_j+E_j$ (Fig.~\ref{fig1}b).

We expand the wave function as $|\psi\rangle = \sum_{j=0}^2c_j|j\rangle$ and introduce $\tilde{c}_j=c_je^{-i\omega_jt}$ ($j=0,1$). In the rotating-wave approximation\cite{separate} (RWA), $\tilde{\psi}=(\tilde{c}_0,\tilde{c}_1,c_2)^\mathrm{T}$ then obeys $id_t\tilde{\psi}=\tilde{H}\tilde{\psi}$ with the time-independent\cite{AFL}
\beq
  \tilde{H}=\begin{pmatrix} \delta_0 & 0 & u_0^* \\ 0 & \delta_1 & u_1^* \\ u_0 & u_1 & 0
                 \end{pmatrix},
\eeql{Heff}
in terms of the only relevant matrix elements $ u_j = \langle2|V_j|j\rangle$. In particular, setting $\omega_0=\omega_1\equiv\omega$ and $V_0=V_1\equiv V$ in regime~(a), (\ref{Heff}) holds in both regimes.

A watershed now occurs between the case of equal detunings\cite{Lu} $\delta_0=\delta_1$, which will shortly be reduced to the standard Rabi problem, and the more complicated $\delta_0\ne\delta_1$, which however does not correspond to a useful gate operation. Note that in regime~(a), the former case is the simple one of degenerate qubit levels; this may be the preferred mode of operation in practice.

First taking $\delta_0=\delta_1=\delta$, we define the Rabi frequency
\beq
  \Omega_\mathrm{R} = \sqrt{\delta^2\!/4 + |u_0^2| + |u_1^2|}\;,
\eeql{OmR}
and a mixing angle $0{<}\eta{<}\pi$ by $\tan\eta=2\sqrt{|u_0^2| + |u_1^2|}/\delta$. One readily finds the inert eigenfunction
\beq
  \tilde{\psi}_0=\frac{(u_1,-u_0,0)^\mathrm{T}}{\Omega_\mathrm{R}\sin\eta}
\eeql{psi0}
obeying $\tilde{H}\tilde{\psi}_0=\tilde{\nu}_0\tilde{\psi}_0$ with $\tilde{\nu}_0=\delta$, which is decoupled from $|2\rangle$ by destructive interference of the microwaves.\cite{shore} In the complementary $2\times2$ space, simple algebra yields the rest of the spectrum as $\tilde{\nu}_\pm=\Omega_\mathrm{R}(\cos\eta\pm1)$,
\begin{align}
  \tilde{\psi}_+&=\left(\frac{u_0^*}{2\Omega_\mathrm{R}\sin\eta/2},
                        \frac{u_1^*}{2\Omega_\mathrm{R}\sin\eta/2},
                        \sin\eta/2\right)^{\!\!\mathrm{T}},\\
  \tilde{\psi}_-&=\left(\frac{u_0^*}{2\Omega_\mathrm{R}\cos\eta/2},
                        \frac{u_1^*}{2\Omega_\mathrm{R}\cos\eta/2},
                        -\cos\eta/2\right)^{\!\!\mathrm{T}}.\label{psipm}
\end{align}

\begin{figure}[t]
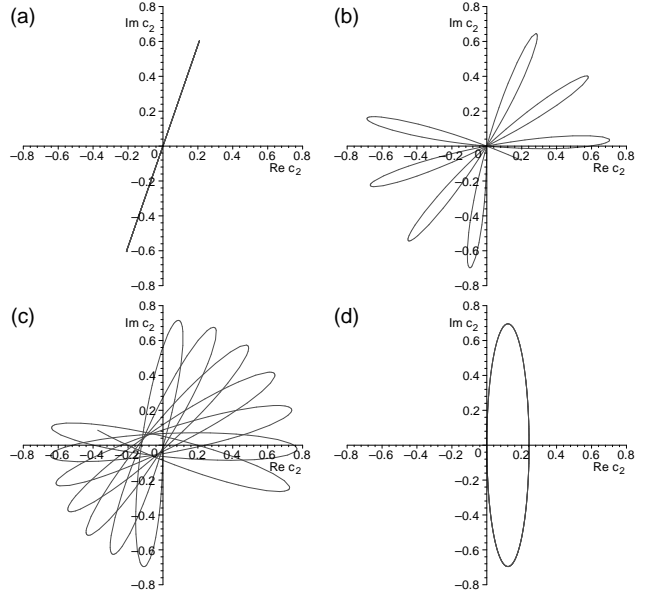

  \includegraphics[width=4cm]{fig2a.eps}
  \kern 2mm\includegraphics[width=4cm]{fig2b.eps}

  \vspace{1mm}
  \includegraphics[width=4cm]{fig2c.eps}
  \kern 2mm\includegraphics[width=4cm]{fig2d.eps}
  \caption{Evolution of $c_2(t)\equiv\langle2|\tilde{U}(t)|0\rangle$ for: (a)~$\delta_0=\delta_1=0$, $u_0=e^{-i/3}$, $u_1=1.2$, $0{\le}t{\le}5\pi$; (b)~$\delta_0=\delta_1=0.25$, $u_0=u_1=\nobreak1$, $0{\le}t{\le}5\pi$; (c)~$\delta_0=0.25$, $\delta_1=0$, $u_0=u_1=1$, $0{\le}t{\le}10\pi$; (d)~$-\delta_0=\delta_1=0.25$, $u_0=1$, $u_1=e^{i/3}$, $0{\le}t{\le}5\pi$.} \label{c2}
\end{figure}

In terms of (\ref{psi0})--(\ref{psipm}), it is trivial to compute the evolution over half a Rabi period $\tilde{U}(t_\mathrm{c})=\exp\{-i\tilde{H}t_\mathrm{c}\}$, decomposing into a reduced gate action $\tilde{U}_\mathrm{r}$ in the logical space and a trivial phase for $|2\rangle$ [cf.\ above (\ref{H}) and Figs.~\ref{c2}a,b]. Only the former concerns us here,
\beq
  \tilde{U}_\mathrm{r}=\frac{1}{\Omega_\mathrm{R}^2\sin^2\!\eta}
  \begin{pmatrix} \zeta|u_0^2|{+}\zeta^2|u_1^2| & u_0^*u_1^\phs(\zeta{-}\zeta^2) \\[1mm]
    u_0^\phs u_1^*(\zeta{-}\zeta^2) & \zeta^2|u_0^2|{+}\zeta|u_1^2| \end{pmatrix},
\eeql{Ur}
a central result, with $\zeta=-e^{-\pi i\cos\eta}$ running through the unit circle with detuning. Clearly, $\tilde{U}_\mathrm{r}$ is unitary, overcoming the problem\cite{han} mentioned in the introduction. The repeated evolution $\tilde{U}(nt_\mathrm{c})_\mathrm{r}=\tilde{U}_\mathrm{r}^n$ follows by simply putting $\zeta\mapsto\zeta^n$ in (\ref{Ur}); hence, the only advantage of taking $n>1$ seems to lie in accessing $\zeta^n\approx1$ without large detuning.

Let us demonstrate that already in its two simplest limits, (\ref{Ur}) is flexible enough to lead to universal computing; contrast Refs.~\onlinecite{han,BHK}. For unbiased systems with symmetric potential [cf.\ (\ref{Usq}), (\ref{Udw}) below] and $u_0=u_1$,
\beq
  \tilde{U}_\mathrm{r,sy}\hn\biggl(\frac{\delta}{\Omega_\mathrm{R}}\biggr)=
  \exp\biggl\{\hn i\biggl[\frac{\pi}{2}-\frac{3\pi\delta}{4\Omega_\mathrm{R}}
  +\biggl(\hn\frac{\pi}{2}{+}\frac{\pi\delta}{4\Omega_\mathrm{R}}\biggr)
  \sigma_{\hn x}\hn\biggr]\biggr\}\;.
\eeql{Urs}
One can also drive at resonance $\delta=0$, but with arbitrary $|u_0/u_1|$ [in regime (b)]. Setting $(|u_0^2|{-}|u_1^2|)/(|u_0^2|{+}|u_1^2|)=\cos\xi$, one has $2u_0^*u_1^\phs/(|u_0^2|{+}|u_1^2|)=e^{i\gamma}\sin\xi$, and
\beq
  \tilde{U}_\mathrm{r,res}(\xi)=ie^{i(\pi/4+\gamma/2)\sigma_{\!z}}
  e^{i\xi\sigma_{\hn x}}e^{i(\pi/4-\gamma/2)\sigma_{\!z}}\;.
\eeql{Urr}
Of course, one always has the phase shifts $e^{i\chi\sigma_{\!z}}$ available, by applying a small bias but no microwaves. Thus, the equivalence above (\ref{H}) is quantitative: adding either (\ref{Urs}) or (\ref{Urr}) suffices to generate all one-bit gates. For $u_0=\nobreak u_1$, $\delta=0$, both of the above reduce to a quantum \textsc{NOT} $\tilde{U}_\mathrm{r}\propto\sigma_{\hn x}$; in general, $[\tilde{U}_\mathrm{r},\sigma_{\hn z}]\ne0$ unless $u_0u_1=0$.

In the ``laboratory frame'' $\psi_\mathrm{r}=(c_0,c_1)^\mathrm{T}$, $U_\mathrm{r}\propto e^{i(\omega_0-\omega_1)t_\mathrm{c}\sigma_{\!z}/2}\tilde{U}_\mathrm{r}= e^{i(E_1-E_0)t_\mathrm{c}\sigma_{\!z}/2}\tilde{U}_\mathrm{r}$ [$=\tilde{U}_\mathrm{r}$ in regime~(a)]. For this specific form, it is assumed that the gate operation starts at $t=0$; this fixes the phases of $V_{0,1}$ in~(\ref{Vt}).

The effective operation rate $\Omega_\mathrm{R}$ in (\ref{OmR}) depends on \emph{intra}well matrix elements $u_j$, between wave functions having an overlap of $O(1)$. For reasonable microwave powers, one thus expects a speedup compared to conventional designs relying on a small $\Delta$. Indeed, the analysis of Ref.~\onlinecite{han} applies, showing that the number of operations achievable in $\tau_\varphi$ is increased by an order of magnitude. If anything, the present situation is slightly more favorable still, since our gate operation is a \emph{one}-step process.

Generalizing the above to $\delta_0\ne\delta_1$ would lead to tedious cubic equations. Fortunately however, this is unnecessary since the crucial decomposition of $\tilde{U}(t)$ then does not generally occur for any finite~$t$.\cite{except} To see this, start e.g.\ from $|0\rangle$ and plot $c_2(t)$ by diagonalizing a few instances of (\ref{Heff}) numerically. The locus of $c_2$ will evolve in a daisy-like pattern (Fig.~\ref{c2}c), without returning to the origin like it does periodically for equal detunings (Figs.~\ref{c2}a,b). These numerics can be supplemented with an expansion in $\delta_0-\delta_1$, the case $\delta_0=\delta_1=0$ being a particularly simple zeroth-order problem.

Some idealizations have been made in the above: $H$~as in (\ref{H}), (\ref{H0}) is a low-dimensional approximation and the effective (\ref{Heff}) follows only in the RWA. The pertinent errors typically are $\sim|V|/|\Delta E|$, where $\Delta E$ can be the distance $E_2-E'$ (positive or negative) to an ignored level $E'$ or $E_1-E_0$ in regime~(b), etc. These can be reduced using a narrow-band, low-power source, but only under the condition $\Omega_\mathrm{R}\tau_\varphi\gg\nobreak1$ of fast gate operation. The issue is well-understood, and techniques such as pulse-shaping exist to counteract off-resonant (including counter-rotating) errors,\cite{correct} in addition to general quantum error-correction methods. The same holds for timing errors.

We now propose two exemplary implementations.

\paragraph*{1.\ SQUIDs.}
One can use any SQUID qubit, such as the three-junction\cite{mooij} or the usual rf\nobreakdash-SQUID. The latter consists of a superconducting ring interrupted by a junction with Josephson energy $E_\mathrm{J}$. The free energy is
\beq
  \mathcal{U}(\phi) = \frac{(\Phi_0\phi/2\pi-\Phi_\mathrm{e})^2}{2L}-E_\mathrm{J} \cos \phi\;,
\eeql{Usq}
with $\phi$ the phase difference across the junction and $\Phi_0=\pi/e$ the flux quantum. When the external flux $\Phi_\mathrm{e}=\Phi_0/2$ and the ring inductance $L > \Phi_0^2/4\pi^2\hn E_\mathrm{J}$, $\mathcal{U}$ will have the bistable shape of Fig.~\ref{fig1}a. The states $|0\rangle$ and $|1\rangle$ correspond to opposite directions of persistent current.

A deviation of $\Phi_\mathrm{e}$ from $\Phi_0/2$ tilts $\mathcal{U}$ (Fig.~\ref{fig1}b), generating a $\sigma_{\hn z}$ operation; applying an rf flux performs a $\sigma_{\hn x}$-like gate (\ref{Ur}). To read out the qubit one should measure the SQUID-generated flux at $\Phi_\mathrm{e}=\Phi_0/2$; its two directions correspond to the logical states.

\paragraph*{2.\ Current-biased $d$-wave junctions.}
In $d$-wave grain boundaries, the order parameter is oriented differently on the two sides of the junction. The resulting Josephson potential is intrinsically bistable,\cite{amin,Ilichev01,refocus,zagoskin} realizing Fig.~\ref{fig1}.

In general, the current--phase relation can have many harmonics. Here, we approximate $I(\phi)=I_1 \sin \phi - I_2 \sin 2 \phi$, where $I$ is the current through and $\phi$ the phase difference across the junction. The free energy thus is
\beq
  \mathcal{U}(\phi) =
    -E_\mathrm{J}\!\left[ \cos \phi -\frac{\alpha}{4} \cos(2 \phi)\right] -
    \frac{I_\mathrm{b}}{2e} \phi\;,
\eeql{Udw}
where $E_\mathrm{J} = I_1/2e$ is the Josephson energy corresponding to the first harmonic, $\alpha = 2 I_2/I_1$, and $I_\mathrm{b}$ is the bias current. When $I_\mathrm{b}=0$, the minima of (\ref{Udw}) are located at
\beq
  \phi = \left\{\!\!\begin{array}{ll}
  \pm \arccos (1/\alpha)\;, & \qquad \alpha > 1\;; \\
  0\;, & \qquad \alpha \le 1\;.\end{array} \right. \label{alpha}
\eeq
For $\alpha > 1$, $\mathcal{U}$ thus is doubly degenerate, with barrier height $\delta\mathcal{U}=E_\mathrm{J}( \alpha  + \alpha^{-1} - 2)/2$ between the minima.

\begin{figure}[t]
\includegraphics[width=40mm]{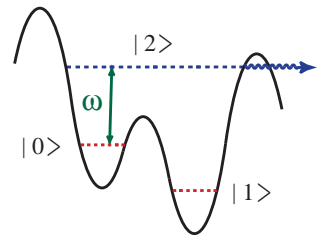}
\caption{Qubit readout using microwave-assisted tunneling to the resistive state. Only the left state will tunnel out.} \label{fig2}
\end{figure}

A finite $I_\mathrm{b}$ removes the degeneracy; this can be used for the $\sigma_{\hn z}$ operation. The gate (\ref{Ur}) can be performed using ac bias currents with appropriate frequencies, as discussed before. For readout, we apply an $I_\mathrm{b}$ such that one of the excited states has a high probability of tunneling to the continuum (Fig.~\ref{fig2}). By selectively coupling one logical state to this excited level, we can determine the qubit state by measuring the junction voltage.\cite{martinis}

Decoherence in $d$-wave qubits is a controversial subject but not central here, so we merely mention a few sources besides external noise (e.g., in $I_\mathrm{b}$). The contribution of ungapped nodal quasiparticles is often overestimated: for a misoriented grain boundary, a node on one side always faces a gapped direction on the other, suppressing tunneling exponentially.\cite{bruder} More problematic are midgap (Andreev) states. Still, since these are split at the qubit's working point, the decoherence due to them can be shown to be tolerable.\cite{golubov}

As a sideline, a classic double-well system with a tunnel splitting is the NH$_3$ molecule. Taking a heavier central nucleus, one arrives at PH$_3$ and AsH$_3$ as instances of Fig.~\ref{fig1}a on a much larger energy scale.\cite{NH3}

In conclusion, it has been shown that microwave coupling via an auxiliary level suffices for coherent control of a Josephson-phase qubit. The advantages include comparative tolerance to device-parameter spread, ability to operate without refocusing, and speed. Charge-noise tolerance (cf.\ the first paragraph) should be excellent: without a need for $\phi$-tunneling, the ratio of $E_\mathrm{J}$ to the charging energy $E_C$ can (and should) be comparatively large. A finite $E_C$ is needed only to ensure appreciable level spacings, as determined by the plasma frequency $\sim\sqrt{E_\mathrm{J}E_C}$; suitable device parameters can be readily chosen. For full-fledged quantum computing, one should additionally describe the coupling of these qubits into a quantum register. While, e.g., tunable-bus proposals\cite{blais} have the promise of being able to couple any type of Josephson qubit, the detailed investigation is still in progress.

We thank A.~Blais and A.M. Zagoskin for their remarks on the manuscript, J.P. Hilton for pointing out the three-pulse sequences in the second paragraph, M.F.H. Steininger for the molecular examples, and T.~Hakio\u glu, J.M. Martinis, and J.~Siewert for discussions.

\pagebreak


\begin{thebibliography}{99}
\bibitem{other} Throughout our paper, $|2\rangle$ need not be the third-lowest level. In fact, in regime (b) [cf.\ above (\ref{Vt})] there can even be states between $|0\rangle$ and $|1\rangle$. However, any such additional states should not be too close to the three indicated ones.
\bibitem{separate} Also, $\langle0|A|1\rangle=0$ for any $A$: the qubit wave functions have exponentially small spatial overlap due to the high tunneling barrier. Without this assumption, if e.g.\ $E_0\approx\nobreak2E_1$ in regime~(b), then the $\omega_1$
pulse could unintenionally irradiate the $|0\rangle\leftrightarrow|1\rangle$ as well
as the $|1\rangle\leftrightarrow|2\rangle$ transition.
\end{thebibliography}
\end{document}